\documentclass[aps,showpacs,nofootinbib]{revtex4}%
\usepackage{amssymb}
\usepackage{graphicx}
\usepackage{amsmath}
\usepackage{amsfonts}%
\newcommand{\ben}{\begin{equation}}
\newcommand{\een}{\end{equation}}
\newcommand{\bes}{\begin{subequations}}
\newcommand{\ees}{\end{subequations}}
\newcommand{\bq}{\begin{eqnarray}}
\newcommand{\eq}{\end{eqnarray}}

\begin{document}
\title{\textbf{\ Lorentz-violating effects on topological defects generated by two real scalar fields}}
\author{D. Bazeia$^{a}$, M. M. Ferreira Jr.$^{b}$, A. R. Gomes$^{c}$ and R.
Menezes$^{a,d}$}
\affiliation{$^{a}${\small {Departamento de F\'{\i}sica, Universidade Federal da Para\'{\i}ba, 58051-970 Jo\~{a}o Pessoa PB, Brazil}}}
\affiliation{$^{b}${\small {Departamento de F\'{\i}sica, Universidade Federal do Maranh\~{a}o,  65085-580 S\~{a}o Luis MA, Brazil}}}
\affiliation{{\small {~}}$^{c}${\small Departamento de F\'{\i}sica, Instituto Federal de Educa\c{c}\~{a}o, C�\^encia e Tecnologia do Maranh\~{a}o,  
65025-001 S\~{a}o Lu\'{\i}s MA, Brazil}}
\affiliation{{\small {~}}$^{d}${\small Departamento de Ci\^encias Exatas, Universidade Federal da Para\'{\i}ba, 58297-000 Rio Tinto PB, Brazil}}

\begin{abstract}
The influence of a Lorentz-violation on soliton solutions generated by a system of two coupled scalar fields is investigated. Lorentz violation is induced by a fixed tensor coefficient that couples the two fields. The Bogomol'nyi method is applied and first-order differential equations are obtained whose solutions minimize energy and are also solutions of the equations of motion. The analysis of the solutions in phase space shows how the stability is modified with the Lorentz violation. It is shown explicitly that the solutions preserve linear stability despite the presence of Lorentz violation. Considering Lorentz violation as a small perturbation, an analytical method is employed to yield analytical solutions.
\\

{Keywords: Lorentz Violation; Domain Wall; Stability.}
\end{abstract}

\pacs{11.30.Cp, 11.30.Er, 03.65.Bz}
\maketitle

\section{Introduction}

Since the demonstration that spontaneous breaking of Lorentz symmetry is
allowed in the high energy context of string theories \cite{Samuel},
Lorentz-violating theories have been extensively studied in\ diverse low
energy systems and used as an effective probe to test the limits of Lorentz
covariance with direct consequences on the Planck scale physics. The
great majority of such investigations take place in the framework of the
Extended Standard Model (ESM), conceived by Colladay and Kosteletck\'{y}
\cite{Colladay}\ as a extension of the minimal Standard Model of the
fundamental interactions. The ESM admits Lorentz and CPT violation in all
sectors of interactions by incorporating tensor terms (generated possibly as
vacuum expectation values of a more fundamental theory) that account for such
a breaking. Actually, the ESM model sets out as an effective model that keeps
unaffected the $SU(3)\times SU(2)\times U(1)$ gauge structure of the
underlying fundamental theory while it breaks Lorentz symmetry at the particle
frame. There is a large amount of work in the literature \cite{Coleman, Electro} searching for experimental
and observational evidence of Lorentz breaking, in an attempt to put lower bound on the parameters.

Topological defects are actually one of the most active research branches of
theoretical physics, with profound connections with condensed matter physics,
gravitation and field theory \cite{topol}. In particular, kink solutions are the simplest topological defects which appear after symmetry breaking of models with one or more scalar fields. Explicit solutions have been worked in several models involving one or more scalar fields \cite{bnrt}, including gravity \cite{grav}, nested defects \cite{nest}, junctions \cite{junct} or connections with other areas of physics \cite{inter}.

The influence of Lorentz violation on kink solutions was investigated in ref. \cite{bbm}, where the first-order framework was applied for some class of models
finding explicit solutions in some interesting cases. Here we pursue this problem studying another class of models with some new results including explicit solutions, phase space analysis, stability analysis and perturbative solutions. This paper is outlined as follows. In Sec. II, we present our model with Lorentz violation and we apply the Bogomol'nyi method to find first-order equations that minimize the energy and solve the equations of motion.  The behavior of numerical solutions are studied in the phase space for several values of the Lorentz breaking parameter. In Sec. III we prove that our solutions are stable and we find that the model support traveling waves. Further, in Sec. IV we find explicit perturbative solutions for the fields. Our final conclusions are presented in Sec. V.

\section{The model and first-order equations}

We consider a class of models with two real scalar fields
\begin{equation}
\mathcal{L}=\frac{1}{2}\partial^{\mu}\phi\partial_{\mu}\phi+\frac{1}%
{2}\partial^{\mu}\chi\partial_{\mu}\chi+k^{\alpha\beta}\partial_{\alpha}%
\phi\partial_{\beta}\chi-V(\phi,\chi), \label{L1}%
\end{equation}
where the coupling tensor $k^{\mu\nu}$ characterizes Lorentz violation. The tensor coefficient is a $2\times2$ matrix,
\[
k^{\alpha\beta}=\left[
\begin{array}
[c]{cc}%
\beta\  & \alpha\\
\alpha & \beta
\end{array}
\right]  ,
\]
and is an element that couples the scalar fields, introducing Lorentz violation and inducing an asymmetry in the propagation of the traveling waves present in these models. The Euler-Lagrange equations can be written as
\bes
\begin{align}
\ddot{\phi}-\phi^{\prime\prime}+\beta(\ddot{\chi}+\chi^{\prime\prime}%
)+2\alpha\dot{\chi}^{\prime}+V_\phi &  =0,\label{D1}\\
\ddot{\chi}-\chi^{\prime\prime}+\beta(\ddot{\phi}+\phi^{\prime\prime}%
)+2\alpha\dot{\phi}^{\prime}+V_\chi &  =0, \label{D2}%
\end{align}
\ees
where $V_\phi=\partial V /\partial\phi$, $V_\chi=\partial V /\partial\chi$, and prime and dot stand for space and time derivative, respectively. In the case of static solutions, we get
\bes
\begin{align}
-\phi^{\prime\prime}+\beta\chi^{\prime\prime}+V_\phi &
=0,\label{E1}\\
-\chi^{\prime\prime}+\beta\phi^{\prime\prime}+V_\chi &  =0. \label{E2}%
\end{align}
\ees
The energy-momentum tensor,
\begin{equation}
T^{\mu\nu}=\frac{1}{2}\left[  \partial^{\mu}\phi\partial^{\nu}\phi
+\partial^{\mu}\chi\partial^{\nu}\chi-k^{\mu\alpha}\partial_{\alpha}%
\chi\partial^{\nu}\phi+k^{\alpha\mu}\partial_{\alpha}\phi\partial^{\nu}%
\chi-g^{\mu\nu}\mathcal{L}\right]  ,
\end{equation}
provides the energy density of the static solutions:
\begin{equation}
T^{00}=\frac{1}{2}\left[  \phi^{\prime2}+\chi^{\prime2}-2\beta\phi^{\prime
}\chi^{\prime}+2V(\phi,\chi)\right]  . \label{T00}%
\end{equation}

A useful form for representing the potential $V(\phi,\chi)$ can be attained from Eqs. (\ref{E1}) and (\ref{E2}), as we get
\begin{equation}
\frac{d}{dx} \biggl( -\frac12{\phi'}^2 -\frac12{\chi'}^2 + \beta\phi'\chi' + V\biggr) = 0.
\end{equation}
Then a potential satisfying
\ben
\label{V}
V=\frac12{\phi'}^2 + \frac12{\chi'}^2 - \beta\phi'\chi'
\end{equation}
would lead to the second-order equations of motion for the fields $\phi$ and $\chi$. This shows that the gradient and potential portions of the energy do not contribute equally, the difference being proportional to the Lorentz violating coefficient $\beta$. 

In order to obtain first-order differential equations whose solutions are also solutions of the equations of motion we use the Bogomol'nyi approach \cite{bog}, looking at the energy density $T^{00}$ and trying to extract simple conditions that minimize the total energy. We introduce a smooth function $W=W(\phi,\chi)$ and consider the {\it ansatz}
\bes\label{A}
\ben
\label{A1}
\phi'=W{_\phi} + s_1\chi' ,
\end{equation}
\ben
\label{A2}
\chi'=W{_\chi} + s_2\phi' .
\end{equation} 
\ees
Substituting this into Eqs. (\ref{T00}) and (\ref{V}) leads to
\begin{eqnarray}
T^{00}&=&\frac{1}{2}(  \phi^{\prime2}+\chi^{\prime2}) -\beta\phi^{\prime
}\chi^{\prime} + \frac12(W_\phi+s_1\chi')^2 + \frac12(W_\chi+s_2\phi')^2 - \beta\phi^{\prime
}\chi^{\prime} \nonumber \\
&=& (\phi'W_\phi + \chi'W\chi) + \phi'\chi'(s_1+s_2-2\beta) + \frac12(\phi'-W_\phi-s_1\chi')^2 + \frac12(\chi'-W_\chi-s_2\phi')^2. \nonumber
\end{eqnarray}
For $s_1+s_2=2\beta$ and when the fields $\phi$ and $\chi$ satisfy the {\it ansatz} given by Eq.~(\ref{A}), we have $T^{00}\equiv\epsilon=dW/dx$, and the total energy $E$ is minimized to the value
\ben
E=\int_{-\infty}^{\infty}  \epsilon(x) dx=|\Delta W|.
\end{equation}
In the remaining of this section we take $s_1=s_2=\beta$.
Explicitly, the {\it ansatz} given by Eq.~(\ref{A}) results in two first-order differential equations that minimize the energy
\bes\label{App}
\begin{eqnarray}\label{A1p}
\phi'&=& \frac{1}{1-\beta^2}(W_\phi+\beta W_\chi)\\
\label{A2p}
\chi'&=& \frac{1}{1-\beta^2}(W_\chi+\beta W_\phi)
\end{eqnarray}
\ees
Note that Eq. (\ref{V}) can be written as 
\ben
V(\phi,\chi)=\frac1{1-\beta^2}\biggl( \frac12W_\phi^2 + \frac12W_\chi^2 + \beta W_\phi W_\chi\biggr),
\end{equation}
and for $\beta^2<1$, which concur with the fact that any Lorentz violating parameter should be very small, the potential is positive. The simplest solutions for the equation \eqref{App} are the homogeneous solution. These solution can be found using the $W_\phi=W_\chi=0$ that are minima of potential.

\subsection{An Example}
Now we will consider a particular potential. We choose the function $W$ introduced in Ref.~\cite{bnrt} 
\ben
\label{W_bnrt}
W(\phi,\chi)=\phi-\frac13\phi^3-r\phi\chi^2
\end{equation}
corresponding to the potential
\ben
V(\phi,\chi)=\frac1{1-\beta^2}\biggl[ \frac12(1-\phi^2-r\chi^2)^2 + 2r^2\phi^2\chi^2 - 2r\beta\phi\chi (1-\phi^2-r\chi^2)\biggr]
\end{equation}
This potential depends explicitly on $\beta$, but the minima are $(\phi,\chi)=(0,\pm\sqrt{1/r})$ and $(\phi,\chi)=(\pm 1,0)$ and they do not depend of $\beta$. These points correspond to the following values of $W$: $P_1=(\phi,\chi)=(-1,0)\Rightarrow W_1=-2/3$; $P_2=(1,0)\Rightarrow W_2=2/3$; $P_3=(0,\sqrt{1/r})\Rightarrow W_3=0$; $P_4=(0,-\sqrt{1/r})\Rightarrow W_4=0$. In this way we can possibly have five BPS sectors connecting $P_1 \leftrightarrow P_2$, $P_1\leftrightarrow P_3$, $P_1\leftrightarrow P_4$, $P_2\leftrightarrow P_3$, $P_2\leftrightarrow P_4$ and one non-BPS sector connecting $P_3\leftrightarrow P_4$.       

Now we analyze Eqs.~(\ref{App}) on the $(\phi,\chi)$ plane, looking for explicit solutions. Those equations reduce, in the model considered here, to the following first-order equations for the fields: 
\bes\label{PCbeta}
\begin{eqnarray}
\label{phibeta}
\phi^\prime&=& \frac1{1-\beta^2}[1-\phi^2-r\chi^2-2\beta r\phi\chi]\\
\label{chibeta}
\chi^\prime&=& \frac1{1-\beta^2}[-2 r\phi\chi+\beta(1-\phi^2-r\chi^2)]
\end{eqnarray}  
\ees
The case $\beta=0$ lead to 
\bes
\begin{eqnarray}
\label{phibeta0}
\phi^\prime &=& 1-\phi^2-r\chi^2, \\ \label{chibeta0}
\chi^\prime &=& -2 r\phi\chi,
\end{eqnarray}  
\ees
which can be integrated to give the general orbit constraint
\ben
\phi^2=\frac{r}{2r-1}\chi^2 + C  \chi^{\frac1r} +1
\een
A special case is the elliptic orbit for $C=0$, with $\phi^2+\frac r{1-2r}\chi^2=1$. As this particular orbit connects the minimum energy points $P_1$ and $P_2$, it is of particular interest with explicit solutions
\ben
\phi(x)=\tanh(2rx),\;\;\;\;\;
\chi(x)=\sqrt{\frac1r-2}\,\,\,\text{sech}(2rx)
\een
Another special orbit is the line $\chi=0$ ($C\to \infty$). In this case, we take $\phi(x)=\tanh(x)$.

\begin{figure}[h!]
\includegraphics[{angle=0,width=12.0cm,height=5cm}]{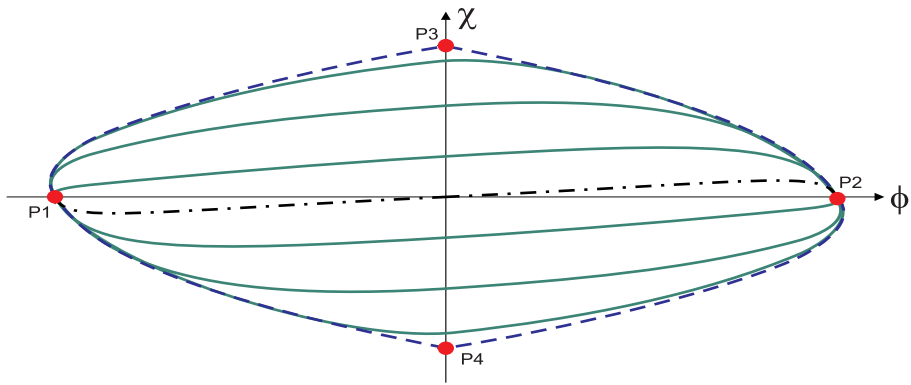} 
\caption{Profile of the orbits in the $(\phi,\chi)$ plane, given by Eqs.~\eqref{PCbeta}, for $\beta=1/3$ and $r=1/3$. The orbit in black (dot-dashed line) corresponds to odd solutions, which have the symmetry $\phi(-x)=-\phi(x)$ and $\chi(-x)=-\chi(x)$.}
\label{orbit}
\end{figure}

We have been unable to find explicit solutions for general nonvanishing values of $\beta$. However, numerically we could obtain orbits and the corresponding solutions $\phi(x)$ and $\chi(x)$. In Fig.~\ref{orbit}, we plot the orbit profile of the Eqs.~\eqref{PCbeta}. The green (solid) lines connect the minima $P_1$ and $P_2$. The blue (dashed) lines connect the minima $P_1$ and $P_2$ to $P_3$ and $P_4$. Note that, in contrast with the case of vanishing $\beta$, there is no solution with one constant field.

\begin{figure}[h!]
\includegraphics[{angle=0,width=5.4cm}]{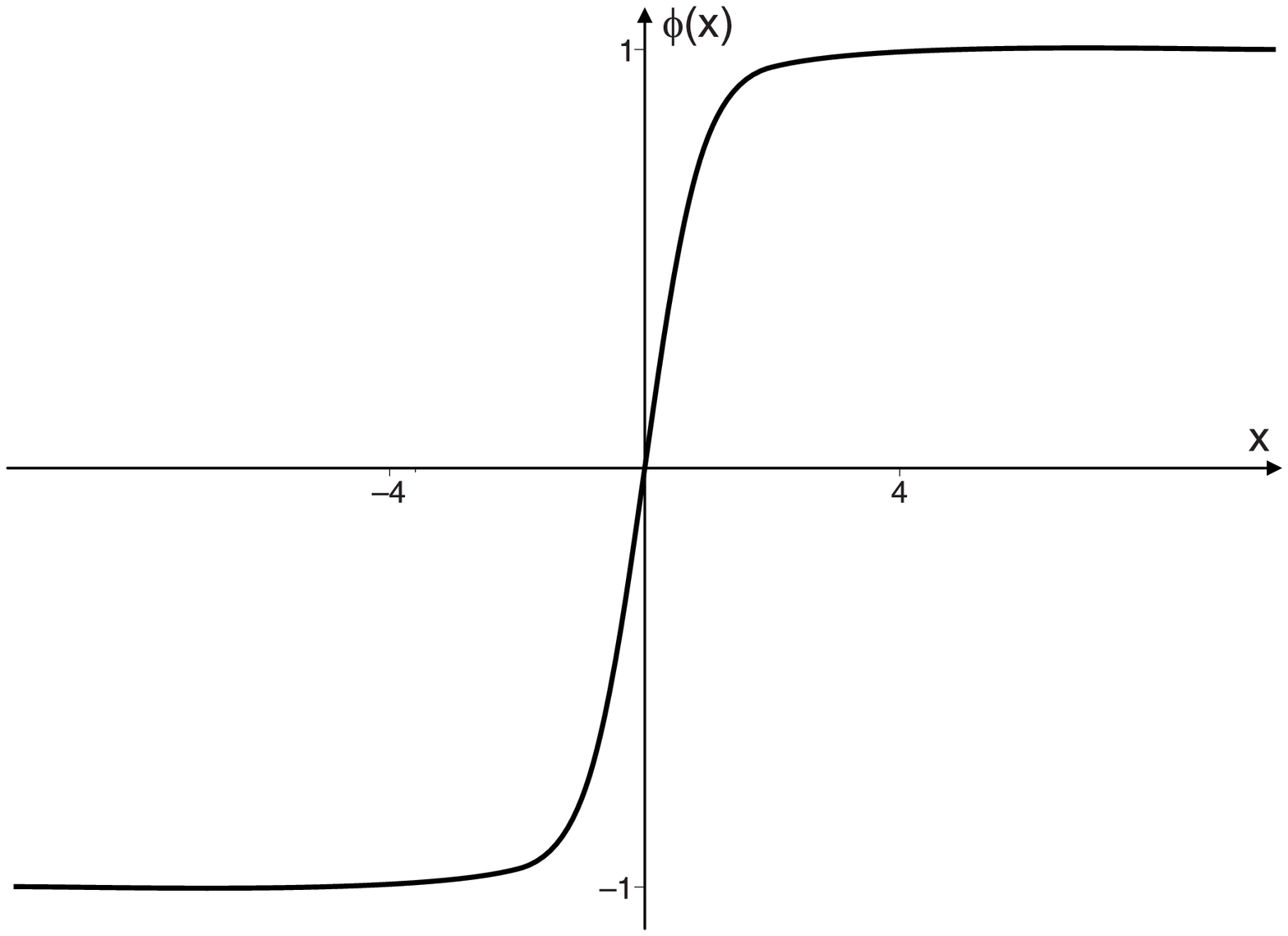} 
\includegraphics[{angle=0,width=5.4cm}]{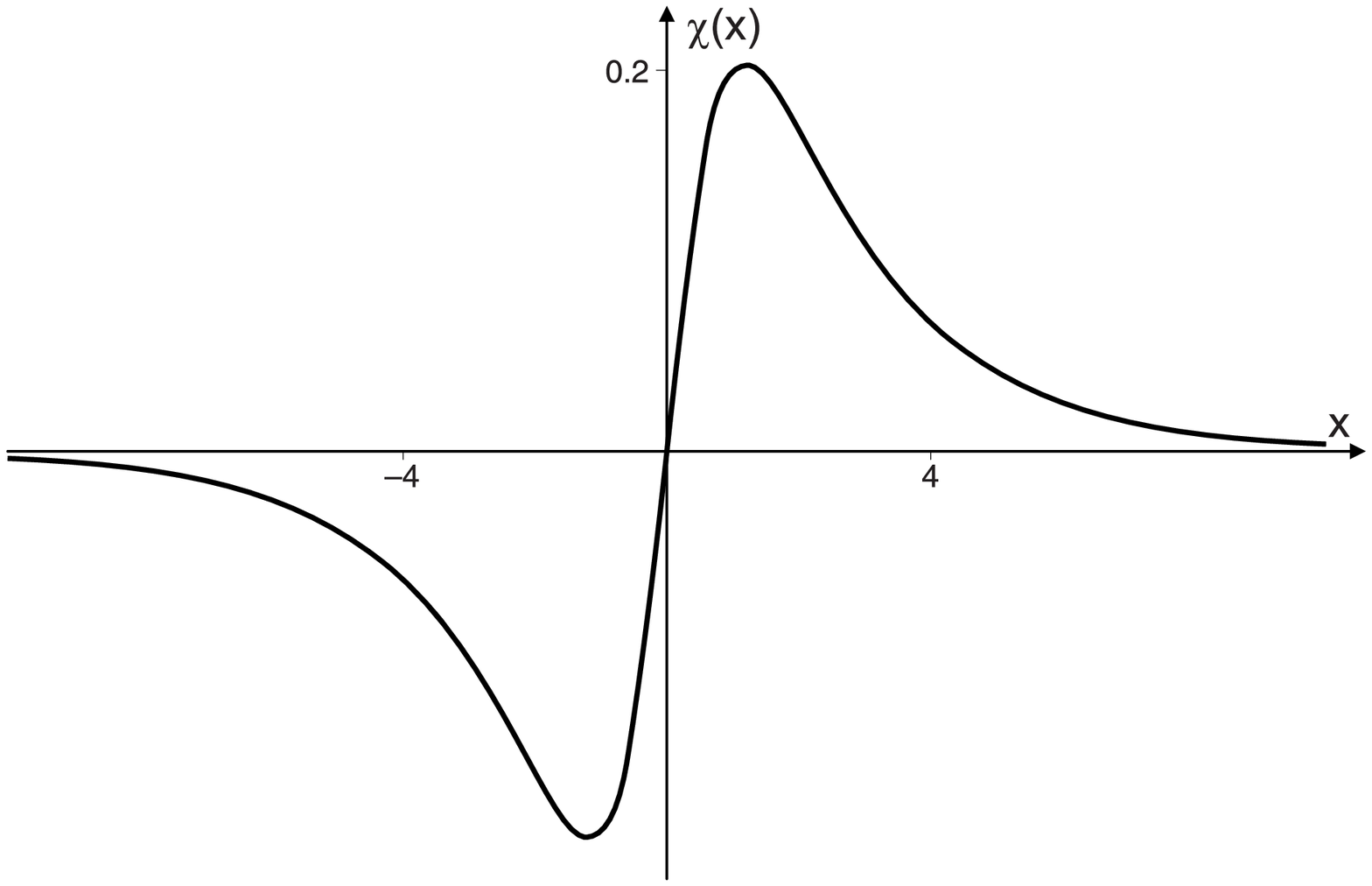}
\includegraphics[{angle=0,width=5.4cm}]{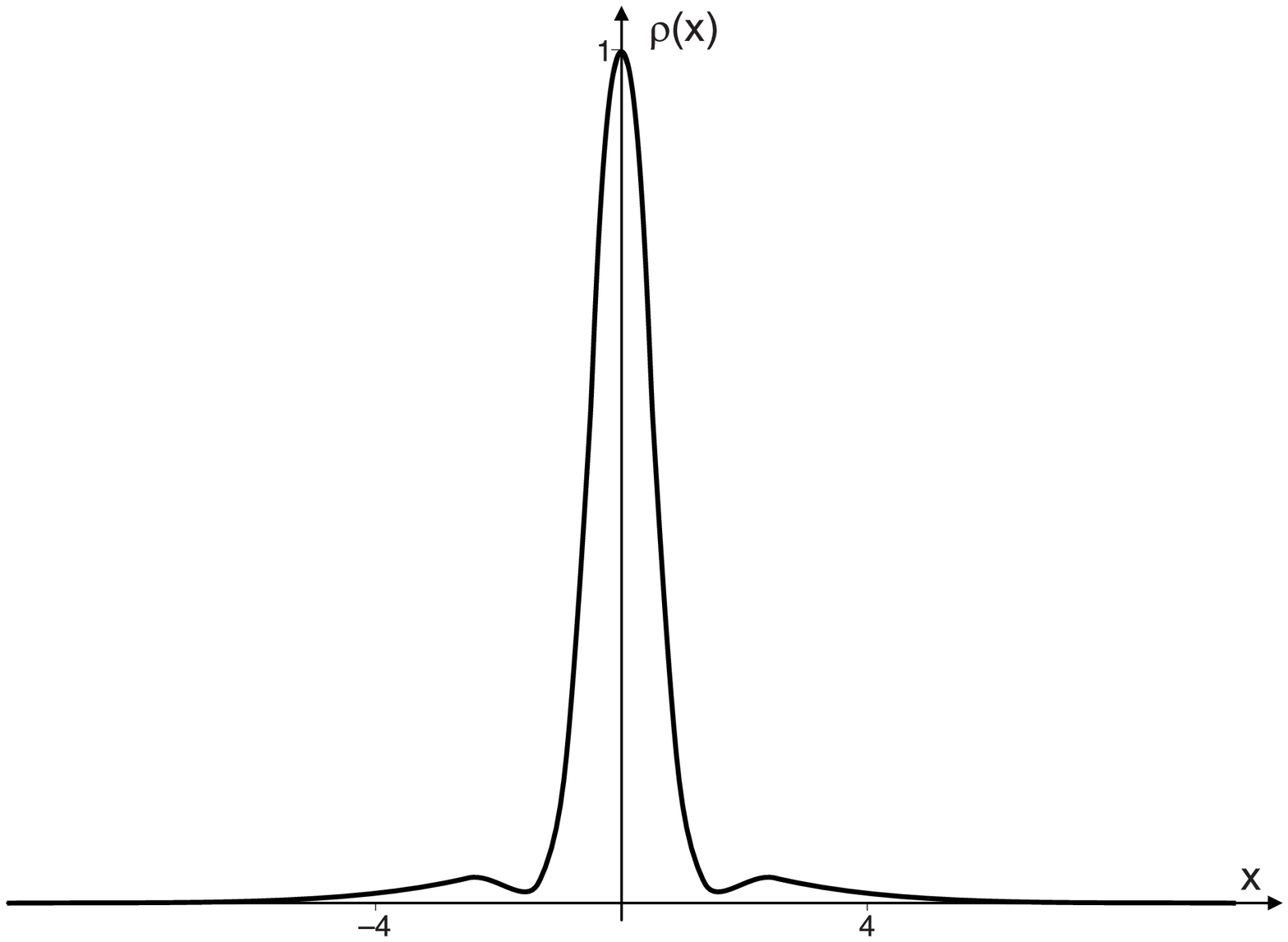}
\caption{The $\phi(x)$ (left panel) and $\chi(x)$ (right panel) solutions and energy density (right panel), for $\beta=1/3$ and $r=1/3$. These solutions  connect the points $P1$ and $P2$.}
\label{phichi1}
\end{figure}

\begin{figure}[h!]
\includegraphics[{angle=0,width=5.2cm}]{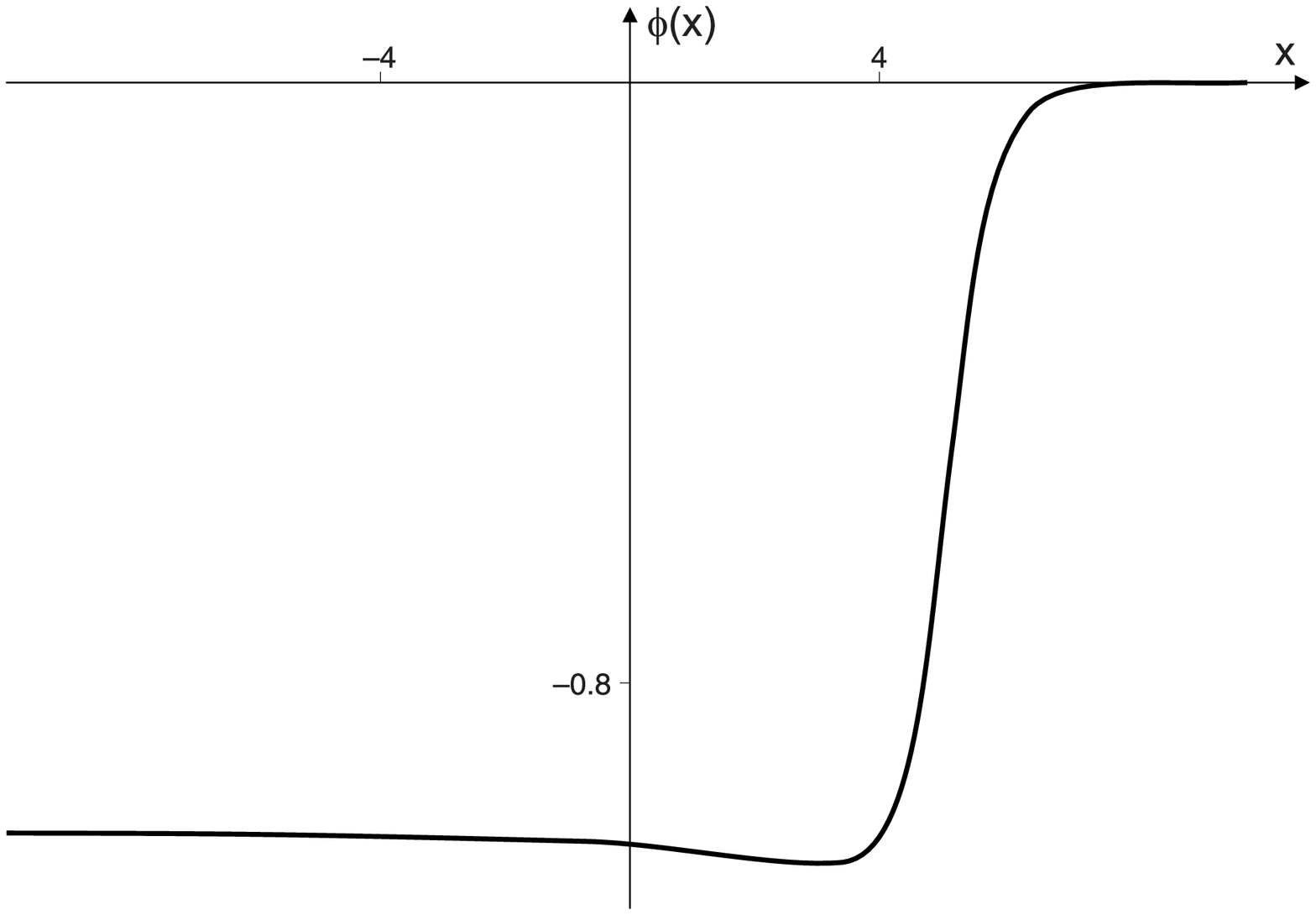} 
\includegraphics[{angle=0,width=5.2cm}]{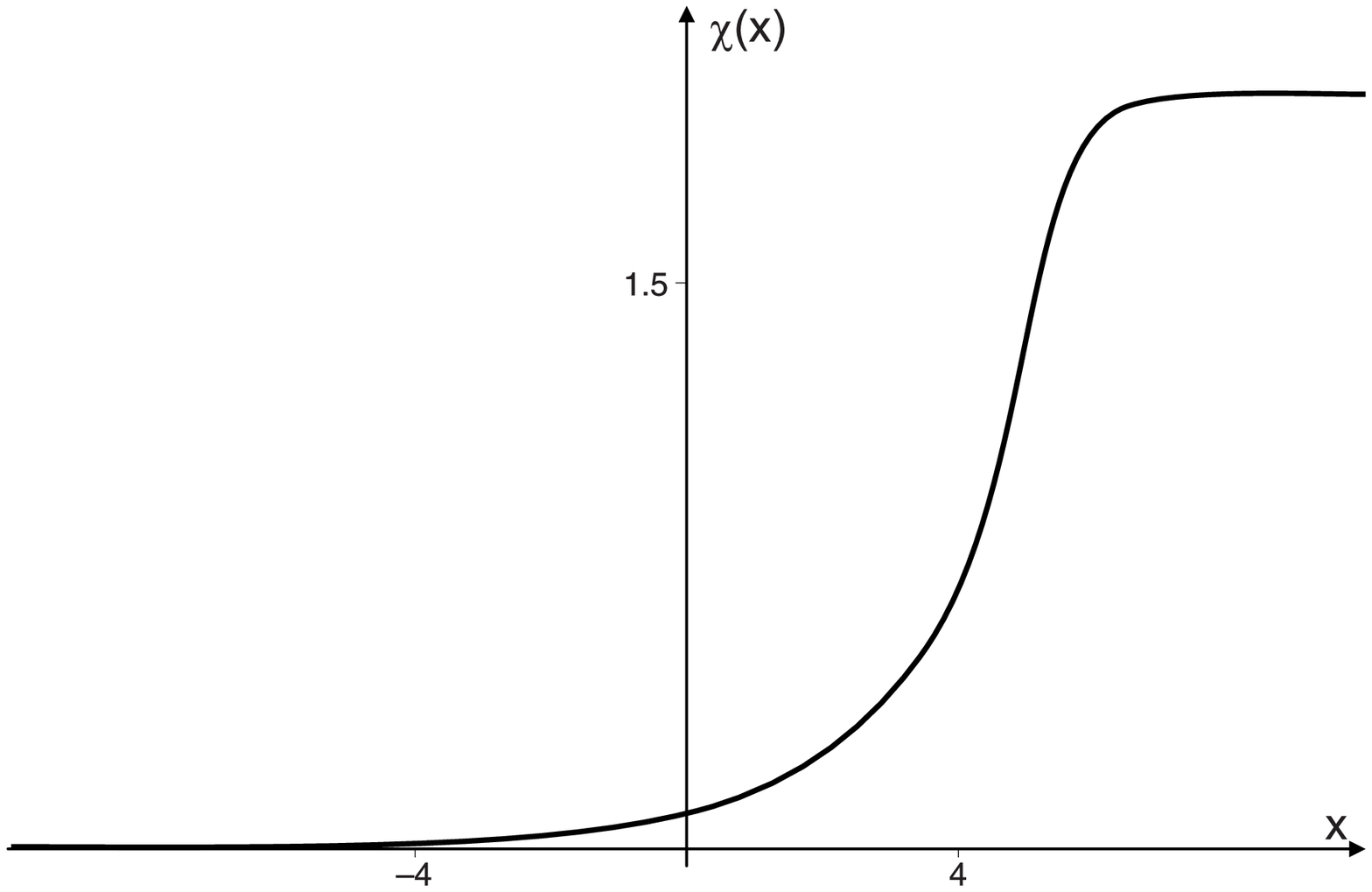}
\includegraphics[{angle=0,width=5.2cm}]{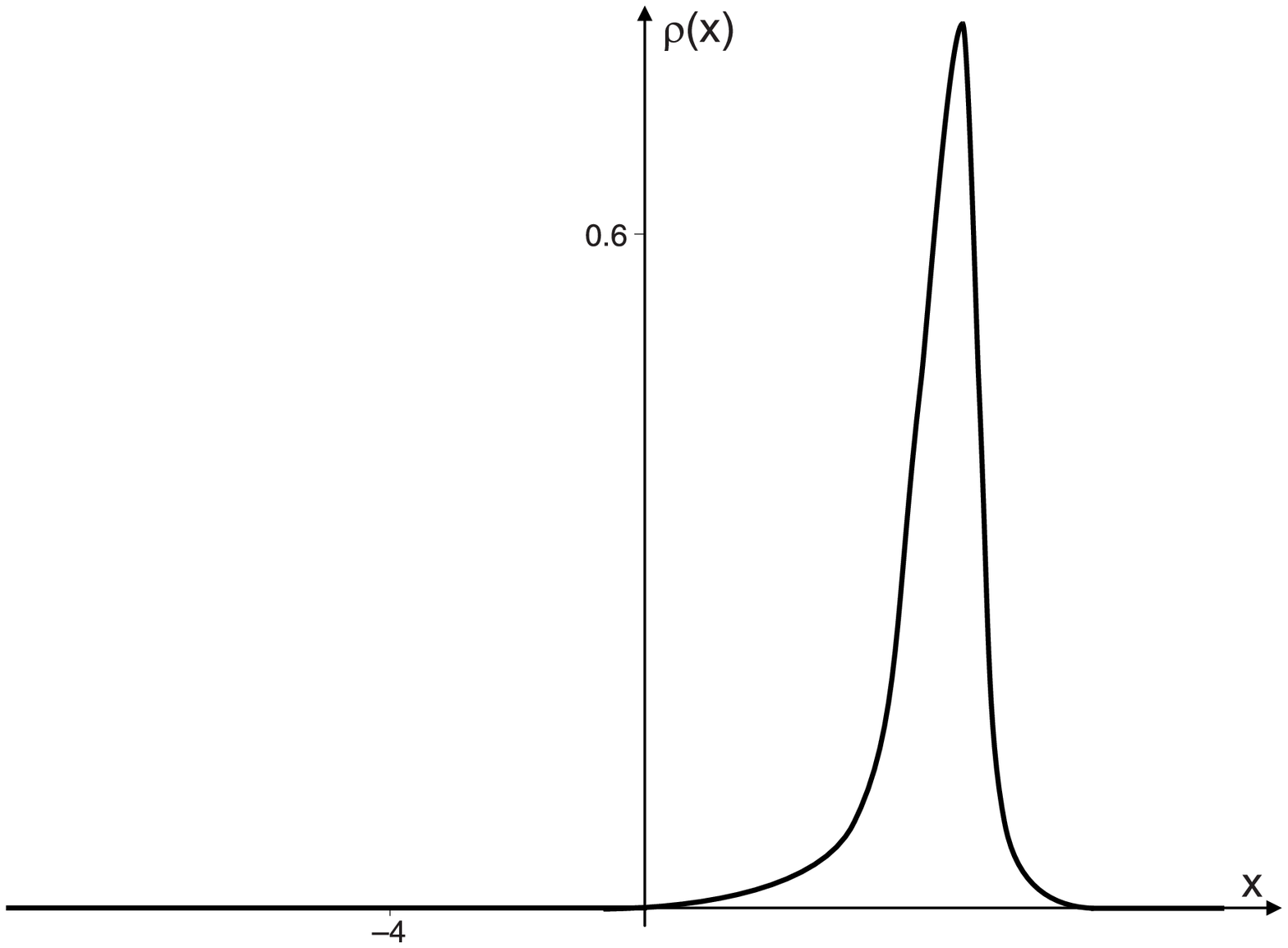}
 \caption{The same as in Fig.~\ref{phichi1}, for solutions that connect the points $P1$ and $P3$.}
\label{phichi2}
\end{figure}

\begin{figure}[h!]
\includegraphics[{angle=0,width=5.2cm}]{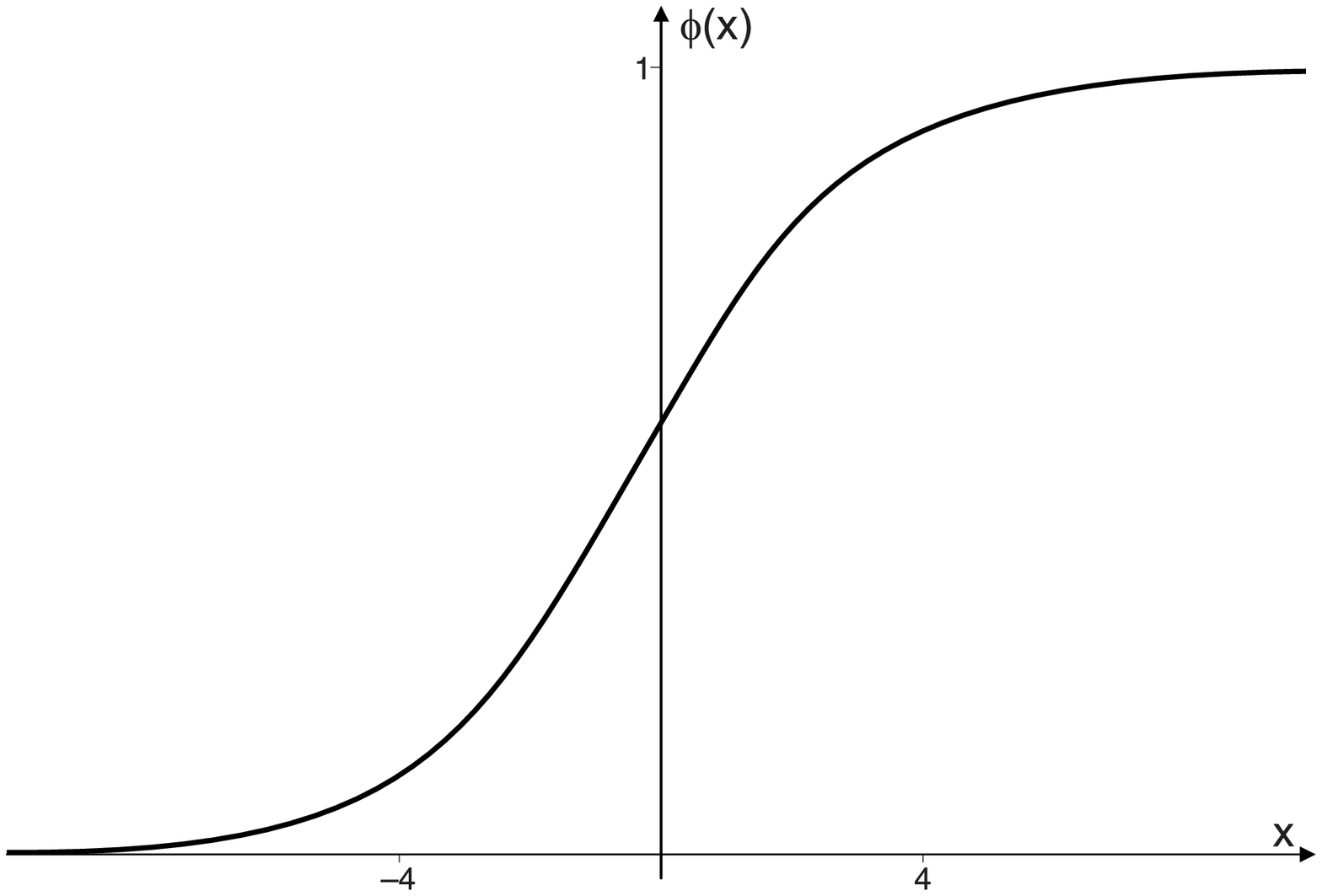} 
\includegraphics[{angle=0,width=5.2cm}]{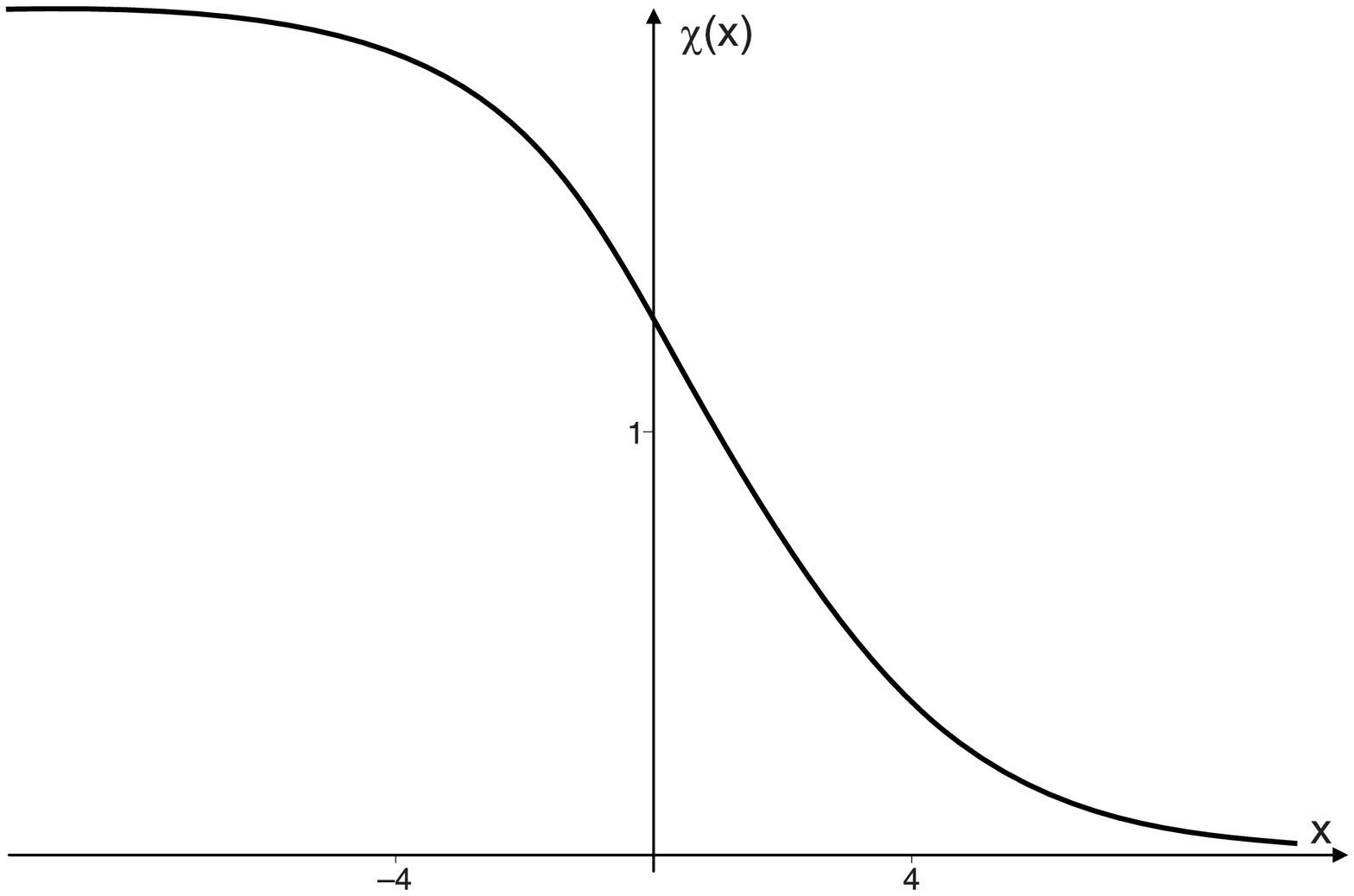}
\includegraphics[{angle=0,width=5.2cm}]{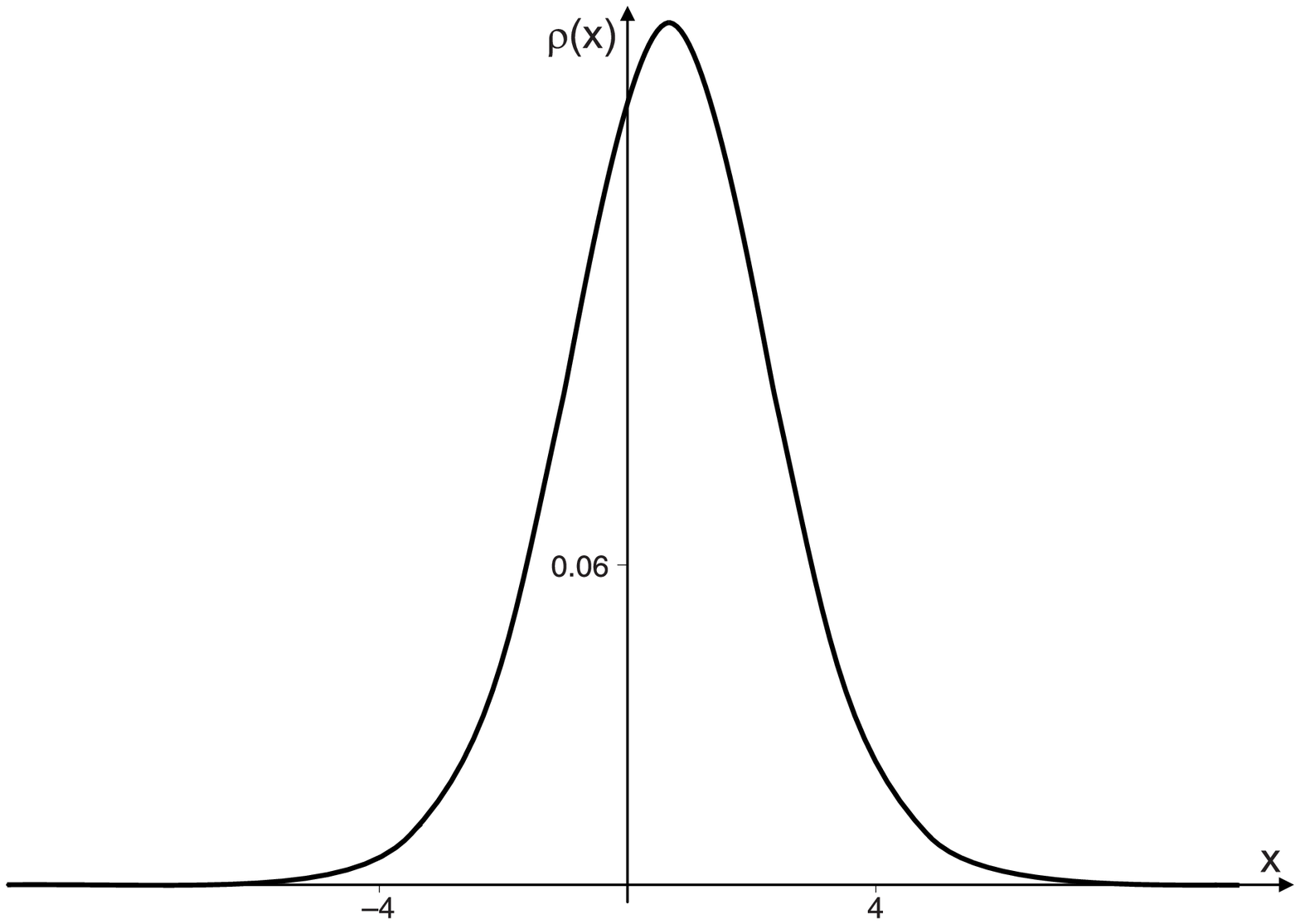}
 \caption{The same as in Fig.~\ref{phichi1}, for solutions that connect $P2$ and $P3$.}
\label{phichi3}
\end{figure}

We plot three pair of solutions in the next three figures, and in Fig.~{\ref{phichi1}} we show that the $\phi(x)$ (left panel) and $\chi(x)$ (center panel) solutions are related to the central orbit highlighted in black in Fig.~\ref{orbit}, which connects $P1$ and $P2$. The right panel shows the energy density, and we note that there are two small energy peaks in addition to the large central peak. These solutions have energy $E_B=4/3$.   
In Fig.~\ref{phichi2}, we show the  solutions and the respective energy density related to the orbit connecting $P1$ and $P3$. The $\phi$ solution is not a monotonic function anymore, with a minimum at $x\approx3.5$. Finally, in Fig.~\ref{phichi3}, we show the solutions that connect $P2$ and $P3$. We note that, in contrast with the $\beta=0$ case, the BPS solutions that connects $P1$ to $P3$ do not have the same form as the BPS solution that connects $P2$ to $P3$, as we can see in Fig.~\ref{phichi2} and in Fig.~\ref{phichi3}. Nevertheless, both solutions have the same energy $E_B=2/3$. 

In the case of $\beta$ being very small, we can find approximate results. However , we will present a perturbative analysis for $\beta\ll1$ in Sect.~\ref{sect_perturb}.

\section{Stability Analysis}

In order to analyze the stability of the solutions, one should start with the full partial differential
equations (\ref{D1}) and (\ref{D2}). For $\alpha=0$, for simplicity, we get to
\bes
\label{EQ12}
\begin{align}
\ddot{\phi}-\phi^{\prime\prime}+\beta(\ddot{\chi}+\chi^{\prime\prime
})+dV/d\phi &  =0,\label{Eq1}\\
\ddot{\chi}-\chi^{\prime\prime}+\beta(\ddot{\phi}+\phi^{\prime\prime
})+dV/d\chi &  =0, \label{Eq2}%
\end{align}
\ees
It is interesting to observe that the Lorentz violation does not jeopardize the linear stability of the differential equations, which turns stable the associated kinks solutions. Such demonstration is performed according to the usual way. With this purpose,
we consider general fluctuations, $\eta,\xi,$ on the field configurations:
\ben
\phi(x,t)=\phi(x)+\eta(x,t),\;\;\;\;\chi(x,t)=\chi(x)+\xi(x,t).
\een
With this, the differential equations (\ref{EQ12}) take the form
\bes
\begin{align}
\alpha_{2}\ddot{\eta}-\alpha_{1}\eta^{\prime\prime}+V_{\varphi_{1}\varphi_{1}%
}\eta+V_{\varphi_{1}\varphi_{2}}\xi &  =0,\label{Dif1}\\
\alpha_{1}\ddot{\xi}-\alpha_{2}\xi^{\prime\prime}+V_{\varphi_{1}\varphi_{2}%
}\eta+V_{\varphi_{2}\varphi_{2}}\xi &  =0. \label{Dif2}%
\end{align}
\ees
where $\alpha_{1}=(1-\beta),\alpha_{2}=(1+\beta)$ and 
\begin{equation}
\varphi_{1}   =\frac{1}{\sqrt{2}}(\phi+\chi),\,\,\,\,\,\,\,\,\,\,\,\varphi_{2}   =\frac{1}{\sqrt{2}}(\phi-\chi), \label{T2}%
\end{equation}

Such system may be written in a matrix form: we use $\eta(x,t)=\eta(x)\cos(\omega t)$ and $\xi(x,t)=\xi(x)\cos(\omega t)$ to get

\begin{equation}
H\left[
\begin{array}
[c]{c}%
\eta\\
\xi
\end{array}
\right]  =\omega^{2}\left[
\begin{array}
[c]{c}%
\alpha_{2}\eta\\
\alpha_{1}\xi
\end{array}
\right]  \label{H1}%
\end{equation}
where
\begin{equation}
H=\left[
\begin{array}
[c]{cc}%
-\alpha_{1}\frac{d^{2}}{dx^{2}}+V_{\varphi_{1}\varphi_{1}} & V_{\varphi
_{1}\varphi_{2}}\\
V_{\varphi_{1}\varphi_{2}} & -\alpha_{2}\frac{d^{2}}{dx^{2}}+V_{\varphi
_{2}\varphi_{2}}%
\end{array}
\right]  .
\end{equation}
Finally, it is possible to show that $H=S^{\dagger}S,$ with:%
\begin{equation}
S=\left[
\begin{array}
[c]{cc}%
-\sqrt{\alpha_{1}}\frac{d}{dx}+p_{1}W_{\varphi_{1}\varphi_{1}} &
p_{2}W_{\varphi_{1}\varphi_{2}}\\
p_{1}W_{\varphi_{1}\varphi_{2}} & -\sqrt{\alpha_{2}}\frac{d}{dx}%
+p_{2}W_{\varphi_{2}\varphi_{2}}%
\end{array}
\right]  .
\end{equation}
with $p_{1}=a\sqrt{\alpha_{1}},$ $p_{2}=b\sqrt{\alpha_{2}}.$ Eq. (\ref{H1}) is
not an eigenvalue equation but its structure is enough to demonstrate stability. Indeed, such equation may be written as $H|\Psi\rangle=\omega_{n}^{2}M|\Psi\rangle,$ where $M$ is a diagonal $2\times2$ matrix with $M_{11}=\alpha_{2}%
,M_{22}=\alpha_{1}.$ Considering the internal product $\langle\Psi
|H|\Psi\rangle=\langle\Psi|S^{\dagger}S|\Psi\rangle=|S|\Psi\rangle|^{2},$ we
obtain $|S|\Psi\rangle|^{2}=$ $\omega_{n}^{2}(\alpha_{2}|\eta|^{2}+\alpha
_{1}|\xi|^{2}),$ which assures positivity of $\omega^{2}$, since $\alpha_{1},\alpha_{2}>0.$ The positivity of $\omega^{2}$ implies
real values for $\omega$, which yields well behaved solutions for any time, meaning stability.

Another study concerns the propagation of traveling waves in this model of two coupled fields. First of all, we note from Eqs.~(\ref{EQ12}) that
\bes
\label{possible}
\begin{align}
\alpha_{2}\ddot{\varphi}_{1}-\alpha_{1}\varphi_{1}^{\prime\prime}%
+dV/d\varphi_{1}  &  =0,\label{phi1}\\
\alpha_{1}\ddot{\varphi}_{2}-\alpha_{2}\varphi_{2}^{\prime\prime}%
+dV/d\varphi_{2}  &  =0, \label{phi2}%
\end{align}
\ees
If such a model entails the presence of traveling waves, then there should exist moving waves endowed with the same form as the
static solutions, which satisfy $\alpha_{1}\varphi_{1}^{\prime\prime}=dV/d\varphi_{1},\alpha_{2}\varphi_{2}^{\prime\prime}=dV/d\varphi_{2}.$
In order to properly examine this issue, the fields are written in the form $\varphi_{1}=\varphi_{1}(u_{1}),\varphi_{2}=\varphi_{2}(u_{2}),$ with
$u_{1}=\gamma_{1}(x-v_{1}t),u_{2}=\gamma_{2}(x-v_{2}t),$ where $\gamma_{1},\gamma_{2}$ being constants and $v_{1},$ and $v_{2}$ represent the propagation
velocities. Replacing $\varphi_{1}(u_{1}),\varphi_{2}(u_{2})$ into eqs.~(\ref{possible}), one gets
\bes
\begin{align}
\alpha_{1}d^{2}\varphi_{1}/du^{2} &  =dV/d\varphi_{1},\label{TW1}\\
\alpha_{2}d^{2}\varphi_{2}/du^{2} &  =dV/d\varphi_{2},\label{TW2}%
\end{align}
\ees
after making the following identification
\begin{equation}
\gamma_{1}=[1-(\alpha_{2}/\alpha_{1})v_{1}^{2}]^{-1/2},\text{ }\gamma
_{2}=[1-(\alpha_{1}/\alpha_{2})v_{2}^{2}]^{-1/2}.
\end{equation}
The attainment of Eqs.~(\ref{TW1},\ref{TW2}) confirms that this model supports traveling waves. The phase velocity $\left(v_{ph}=\omega/k\right)$ of such waves is given by the dispersion relations coming from the free version of Eqs.~(\ref{phi1},\ref{phi2}), $-\alpha_{2}k^{2}+\alpha_{1}%
\omega^{2}=0,$ $-\alpha_{1}k^{2}+\alpha_{2}\omega^{2}=0$, which implies $v_{1}=\pm\sqrt{\alpha_{2}/\alpha_{1}},$ $v_{2}=\pm\sqrt{\alpha_{1}/\alpha
_{2}}.$ Considering the redefinitions (\ref{T2}), the stationary wave configurations associated with the fields $\phi,\chi$ are composed by two
travelling waves of velocities $v_{1},$ $v_{2}.$ It is interesting to note that $v_{2}<1$ while $v_{1}>1$, and that this outcome prevails for the group
velocity as well, with $v_{g_{2}}<1$ while $v_{g_{1}}>1.$

\section{Analytical method for perturbed solutions}
\label{sect_perturb}

Taking into account that Lorentz violation is a small effect, a perturbative approach is well suited to deal with it. In this sense, an analytical method
may be used to construct explicit first-order solutions for the differential equations (\ref{EQ12}) constrained to the situation where the
LV parameter $\beta$ is small (since its presence is addressed as a perturbation on the Lorentz symmetric case). In order to obtain first order
analytical solutions, we should first define the unperturbed system $(\beta=0)$ as
\bes
\begin{align}
\ddot{\phi}-\phi^{\prime\prime}+dV/d\phi &=0,\\
\ddot{\chi}-\chi^{\prime\prime}+dV/d\chi &=0,
\end{align}
\ees
which provides unperturbed solutions (labeled as $\phi_{0}$ and $\chi_{0})$, whose form depend on the adopted potential $V.$ In this section we consider $V(\phi,\chi)= V_1(\phi)+V_2(\chi),$ which makes the above equations uncoupled.

In the static perturbed case, the differential equations are:
\bes
\begin{align}
-\phi^{\prime\prime}+\beta\chi^{\prime\prime}+dV/d\phi &  =0,\label{AE1}\\
-\chi^{\prime\prime}+\beta\phi^{\prime\prime}+dV/d\chi &  =0. \label{AE2}%
\end{align}
\ees
Note that these equations are coupled only by $\beta$-dependent terms. In order to solve it, we propose the following ansatz
\ben
\phi(x)=\phi_{0}(x)+\beta g(x),\;\;\;\;\;\chi(x)=\chi_{0}(x)+\beta f(x), \label{Qui3B}%
\een
where $g$ and $f$ are functions to be determined. Replacing such ansatz in
eqs.~(\ref{AE1},\ref{AE2}), it results
\bes
\begin{align}
\label{eqg}
-g^{\prime\prime}+\chi_{0}^{\prime\prime}+gd^{2}V/d\phi^{2}|_{\phi_{0}}&=0,\\
\label{eqf}
-f^{\prime\prime}+\phi_{0}^{\prime\prime}+fd^{2}V/d\chi^{2}|_{\chi_{0}}&=0,
\end{align}
\ees
where $\phi_{0}$ and $\chi_{0}$ are the corresponding unperturbed solutions. These are second-order differential equations, whose solutions may
be achieved in an analytical form for certain potentials. Aa a first example, let us consider the potential such that
\bes
\begin{eqnarray}
\label{Pot}
V_1(\phi)=(1-\phi^{2})^{2}/2,\;\;\;\;\;V_2(\chi)=(1-\chi^{2})^{2}/2,
\end{eqnarray}
\ees
which provides the well-known solutions, $\phi_{0}(x)=\pm\tanh(x),$ $\chi_{0}(x)=\pm\tanh(x),$ besides the homogeneous ones: $\phi_{0}(x)=\chi_{0}(x)=\pm1,0.$ In this way, Eqs.~(\ref{eqg})-(\ref{eqf}) give
\bes
\begin{align}
-g^{\prime\prime}+\chi_{0}^{\prime\prime}+g(-2+6\phi_{0}^{2})  &  =0,\\
-f^{\prime\prime}+\phi_{0}^{\prime\prime}+f(-2+6\chi_{0}^{2})  &  =0.
\end{align}
\ees

A first situation consists in choosing $\phi_{0}(x)=\pm\tanh(x),$ $\chi_{0}(x)=\pm1,$ which leads to the system of equations
\bes
\begin{align}
-g^{\prime\prime}+g(6\tanh^{2}x-2)  &  =0,\label{g2A}\\
-f^{\prime\prime}\mp2\tanh x\operatorname*{sech}\nolimits^{2}x+4f  &
=0. \label{f2A}%
\end{align}
\ees
Under the boundary conditions $(dg/dx)|_{x\to\infty}=(df/dx)|_{x\to\infty}=0,$ $g(0)=a,$ $f(0)=b,$ such system supports an exact analytical solution given
as follows
\bes
\begin{align}
g(x)&=a\operatorname*{sech}\nolimits^{2}x,\label{S1A}\\
f(x)&=be^{-2x}\pm2[1-\ln(2\cosh x)]\sinh2x\pm2x\cosh2x\mp\tanh x(1+2\cosh2x). \label{S1B}
\end{align}
\ees	
With this, the classical solutions for the Lorentz-violating system are determined at first order in accordance with eqs.~(\ref{Qui3B}).

A second case is $\phi_{0}(x)=\pm\tanh(x),$ $\chi_{0}(x)=\tanh(x)$ with the conditions $(dg/dx)|_{x\to\infty}=(df/dx)|_{x\to\infty}=0,$ $g(0)=a,$ $f(0)=b,$
which gives
\ben
g(x)=(a+x/2)\operatorname*{sech}\nolimits^{2}x,\;\;\;\;\;f(x)=(b\pm x/2)\operatorname*{sech}\nolimits^{2}x. \label{S2B}
\een
Once we have obtained first-order analytical solutions, it is also possible to carry out the corresponding first-order energy corrections for the solutions
already known. The starting point is the energy density given by eq.~(\ref{T00}), written for eqs.~(\ref{Qui3B}), which implies
$T^{00}=T_{0}^{00}+T_{\beta}^{00},$ where:
\bes
\begin{align}
T^{00}  &  =\frac{1}{2}\left[  \phi_{0}^{\prime2}+\chi_{0}^{\prime2}%
+2V(\phi,\chi)\right] \\
T_{\beta}^{00}  &  =\beta\lbrack\phi_{0}^{\prime}g^{\prime}+\chi_{0}^{\prime
}f^{\prime}-\phi_{0}^{\prime}\chi_{0}^{\prime}+V_{\phi}g+V_{\chi}f].
\end{align}
\ees
Here, $T_{0}^{00}$ is the usual energy density (without Lorentz-violation) and $T_{\beta}^{00}$ is the first-order correction due to Lorentz violation. Now,
using the potentials of Eqs.~(\ref{Pot})a-(\ref{Pot})b, the energy correction takes the form
\begin{equation}
T_{\beta}^{00}=\beta\lbrack\phi_{0}^{\prime}g^{\prime}+\chi_{0}^{\prime
}f^{\prime}-\phi_{0}^{\prime}\chi_{0}^{\prime}-2\phi_{0}(1-\phi_{0}%
^{2})g-2\chi_{0}(1-\chi_{0}^{2})f].
\end{equation}
A first case whose energy is to be analyzed is the one in which $\phi_{0}(x)=\pm\tanh(x),$ $\chi_{0}(x)=\pm1.$ In this case, the corresponding
solutions are given by eqs.~(\ref{S1A},\ref{S1B}), which lead to the following result
\bes
\begin{align}
T_{0}^{00}  &  =\operatorname*{sech}\nolimits^{2}x,\\
T_{\beta}^{00}  &  =\mp\beta\lbrack6a\tanh x\operatorname*{sech}%
\nolimits^{4}x].
\end{align}
\ees
The energy associated with this density, $E=\int_{-\infty}^{\infty}T_{\beta}^{00}dx,$ is zero, since one has to integrate an odd function. Hence,
the energy is simply the one of the usual case (without LV): $E=\int_{-\infty}^{\infty}T_{0}^{00}dx=4/3.$ This means that the energy does not change up to first-order in $\beta$.

The second case is given by $\phi_{0}(x)=\pm\tanh(x)$ and $\chi_{0}(x)=\tanh(x),$ whose solutions are given by eqs.~(\ref{S2A},\ref{S2B}). In this case, one gets
\bes
\begin{align}
T_{0}^{00}&=2\operatorname*{sech}\nolimits^{2}x,\\
T_{\beta}^{00}&=\mp\frac{4\beta}{3}[x\tanh x\operatorname*{sech}
\nolimits^{4}x].
\end{align}
\ees
The associated energy correction is then equal to $\mp4\beta/3,$ with total energy being given by $E=4(2\mp\beta)/3.$ Here the energy receives first-order correction in $\beta$.  

\section{Ending comments}

In this work, we have studied the effects of a Lorentz-violating tensor term on the topological defects associated with models described by two real scalar fields. We applied the Bogomol'nyi approach to successfully attain first-order differential equations whose solutions minimize the energy and solve the equations of motion. This simplification is extremely useful to study numerically the solutions in phase space. We studied a model where the potential describing the two scalar fields presents nonlinear couplings between the fields, with a rich structure of minima. As we have shown, for general values of the parameter which implements the Lorentz breaking, a numerical investigation was performed, and we have found numerical solutions connecting minima of the potential. For general models, we performed a stability analysis to show that the solutions of the equations of motion are well behaved in time. Also, we have shown that the models support traveling waves in general. Since we where not able to find explicit solutions for general values of $\beta$, we studied the case with $\beta\ll1$, which concur with the fact that the Lorentz breaking should be very small. In this case, we proposed an ansatz for the perturbed solutions, finding explicit solutions for the special case where the potential is described by the two fields $\phi$ and $\chi$, that do not interact with each other.

\begin{acknowledgments}
The authors express their gratitude to CAPES, CNPq, FAPEMA and PADCT-MCT-CNPq for financial support.
\end{acknowledgments}

\end{document}